\documentclass[10pt,twoside]{hsqcd}
\usepackage{epsf,amsmath}
\usepackage{graphicx}
\usepackage{axodraw}
\usepackage{bm}

\begin{document}

\title{Soft gluons are heavy and rowdy
\thanks{Supported by grants FPA 2004-02602, 2005-02327,
DE-FG02-97ER41048, PR27/05-13955-BSCH, DFG FI 970/2-1 and DFG 
FI 970/7-1.}}

\author{Reinhard Alkofer$^1$, Pedro Bicudo $^2$, Stephen R. Cotanch $^3$,\\ 
Christian S. Fischer $^4$ and \underline{Felipe J. Llanes-Estrada}$^5$. \\ \\
$^1 $ Inst. f\"ur Physik, Karl-Franzens Univ. 1-8010 Graz, Austria.\\
$^2 $ Instituto Superior Tecnico, Av. Rovisco Pais, 1049-001 Lisboa, Portugal.\\
$^3 $ Physics Department, North Carolina State University,
Raleigh 27695 NC USA.\\
$^4 $ Gesellschaft f\"ur Schwerionenforschung, Darmstadt, Germany. \\
$^5$ Fisica Teorica I, Universidad Complutense de Madrid, 
28040 Madrid Spain, \\ E-mail: fllanes@fis.ucm.es }

\maketitle

\begin{abstract}
\noindent   We study dynamical mass generation in pure Yang-Mills theory
and report on a recently developed ansatz that exactly solves the tower
of Dyson-Schwinger equations in Landau gauge at low Euclidean momentum,
featuring enhanced
gluon-gluon vertices, a finite ghost-gluon vertex in agreement with an 
old argument of Taylor, and an IR suppressed gluon propagator. This ansatz
reinforces arguments in favor of the concept of a gluon mass gap at low momentum
(although the minimum of the gluon's dispersion relation is not at zero
momentum).
As an application, we have computed the spectrum of oddballs,
three-gluon glueballs with negative parity and C-parity.  The three body problem
is variationally solved employing the color density-density
interaction of Coulomb gauge QCD with a static Cornell potential.
Like their even
glueball counterparts, oddballs fall on Regge trajectories with similar 
slope
to the pomeron. However their intercept at $t=0$ is smaller than the $\omega$
Regge trajectory and therefore the odderon may only be
visible in experimental searches (for example at BNL) with higher $-t$
than conducted to date   at DESY.
\end{abstract}

\markboth{\large \sl R. Alkofer, P. Bicudo, S. Cotanch, C. Fischer and
\underline{F. Llanes-Estrada}} 
{\large \sl \hspace*{1cm} SOFT GLUONS ARE HEAVY AND ROWDY}

\newpage

In a free Yang-Mills theory gauge boson mass terms break gauge invariance
and cannot be part of the Lagrangian. Moreover, if self-interactions are
incorporated
perturbatively, to any finite order in perturbation theory the radiative
corrections do not generate mass terms.
However non-perturbative effects can potentially  provide dynamical suppression
of the gauge boson propagator via a dressing function
\begin{equation}
\frac{1}{p^2} \to \frac{Z(p^2)}{p^2} \ .
\end{equation}
If this function $Z$ is proportional to $p^2$ in the limit $p\to 0$ then
the complete propagator behaves as that of a Yukawa exchange of finite mass.
This behavior was conjectured by Cornwall  \cite{Cornwall:1981zr}. 
Dynamical suppression in  gluodynamics, unlike the electroweak Lagrangian, 
is not facilitated by coupling to an external Anderson-Higgs field, but 
can only be an intrinsic phenomenon due to the non-linearity and strength 
of the self-coupling.

The fact that the positions of resonances in the hadron spectrum seem 
to follow a discrete pattern  is a clear indication that the 
gluon degrees of freedom entail a mass gap of order 800 MeV.

This has been previously incorporated in phenomenological 
models using a constituent gluon mass \cite{Kaidalov:1999yd,Brau:2004xw},
a waveguide cut-off frequency such as the bag model \cite{Barnes:1981kp},
or a quantum string excitation in the flux tube model
\cite{Isgur:1984bm}.

In the last years the suppression of the gluon propagator has been put
on firm theoretical footing  in the framework of the Yang-Mills Dyson-Schwinger
equations 
\cite{Alkofer:2000wg,Alkofer:2004it,Fischer:2005wx,Fischerthesis}, 
and has also received support 
from numerical lattice calculations \cite{Oliveira:2004gy,Bowman,Lokhov:2005bv}.

The studies in \cite{Alkofer:2000wg,Fischer:2005wx,Fischerthesis}, 
employing sophisticated ans\"atze
for the interaction vertices were able to provide analytical and numerical
solutions for the infrared behavior of the Yang-Mills ghost and gluon
propagators. The novelty was that the dressing function for the gluon propagator
in Landau gauge was more suppressed than  $p^2$, leading to an infinite
gluon mass at zero momentum. Conversely, the dressing function for the
ghost propagator turned out to be infrared singular instead of suppressed. In
terms of a real constant $\kappa \in (0.5,0.6)$ (the calculation of \cite{Fischerthesis}
yields 0.596), they read
\begin{equation}\label{propagators}
\lim_{p\to 0} Z(p^2) = {\rm ct \cdot} (p^2)^{2\kappa} \ \
\lim_{p\to 0} G(p^2) = {\rm ct \cdot} (p^2)^{-\kappa}\ .
\end{equation}
In \cite{Alkofer:2004it} further progress has been made by observing that
one can make a self-consistent ansatz that simultaneously solves all 
of the Yang-Mills wave equations  in the far infrared. The result for
the dressing of the Green's function with $m$ gluon legs and $2n$ 
ghost legs, $A[m,2n]$, is
\begin{equation} \label{result}
\lim_{{\rm all} p\to 0} A[m,2n](p^2)  =  {\rm ct}\cdot (p^2)^{(n-m)\kappa} 
\ .
\end{equation}
 By ``dressing'' we understand that the canonical dimension of the Green's
function is still multiplying the result, and by ``all'' that all the momenta
entering the scattering amplitude need to tend to zero simultaneously.
A proof employing the skeleton expansion has been given in \cite{Alkofer:2004it}
and \cite{Fischer:2005wx}. In addition to the aforementioned gluon suppression
and ghost enhancement, one finds from eq. (\ref{result}) that the ghost-gluon
vertex $(n=1,m=1)$ is IR finite in agreement with an argument of Taylor
\cite{Dixon:1974ss} and that the three and four-gluon vertices are enhanced by 
$(p^2)^{-3\kappa}$ and $(p^2)^{-4\kappa}$ dressings respectively. These
relations between the dressings force cancellations such that all the effective
charges defined from the primitively divergent  vertices in the Lagrangian
are finite in the IR, and  $\alpha_s$  presents an IR fixed point, which
theoretically supports approaches based on conformal theory \cite{Brodsky:2005en}
and is compatible with analytical perturbation theory 
\cite{Shirkov:2001hj,Bakulev:2005gw}.

The demonstration that eq. (\ref{result}) indeed solves the tower of 
DSE's has been given in terms of the skeleton expansion. 
Here we give an argument that is independent of this expansion, sidestepping
convergence issues. 

We will examine in turn all the contributions to 
the RHS of the Dyson Schwinger equation for $A[m,2n]$, assume eq. (\ref{result})
for the coupled-channel Green's functions appearing there, and show
that the leading IR divergence yields again our counting rule eq. (\ref{result}).  

There are several formulations of the Dyson-Schwinger equations, we
employ a very intuitive one based on a complete resummation of the 
perturbative series (without the intermediate skeleton expansion), 
that can conceivably be generalized by analytical continuation, and one
should remember that the DSE's can be directly formulated as identities
in the path-integral formalism.

We will draw a box for the Green's function with $m$ gluon and $n$ ghost
legs labeled by the indices $m,2n$. Its DSE reads
\begin{equation}\label{dse}
\begin{picture}(0,0)(200,50)
\Boxc(50,25)(50,25)\put(35,25){m,2n} \put(100,25){$= {\rm Tree}
+(2a)+(2b)+(3)\delta(m)$\ .}
\end{picture}
\end{equation} 
\vspace{1cm}

The tree-level diagrams are the lowest order in perturbation theory and have their
canonical dimension. All others involve loop integrals and can feature
an anomalous infrared  exponent. 
Diagrams of type (2a) have at least one external gluon leg coupled to 
a vertex that is not attached to any other external leg. These diagrams
can be resummed to all orders to one of the three following classes
(we write below the corresponding infrared exponents). All propagators
are understood as fully dressed and carrying their dimensions given by
eq. (\ref{propagators}).
\vspace{-.5cm}

\begin{equation} \nonumber
\begin{picture}(-50,0)(200,100)
\Boxc(25,40)(50,26)   \Boxc(100,40)(50,26) \Boxc(175,40)(50,26)
\put(5,42){$m-1,$} \put(5,32){$2(n+1)$}
 \put(80,38){$m+1,2n$} \put(155,38){$m+2,2n$}
\put(5,0){$\kappa(n-m)$} \put(75,0){$\kappa(n-m+3)$}\put(150,0){$\kappa(n-m+4)$}
\DashLine(2,53)(25,75){4} \DashLine(48,53)(25,75){4} 
\Gluon(25,75)(25,100){3}{3} \Gluon(100,75)(100,100){3}{3}
\Gluon(175,75)(175,100){3}{3}
\Gluon(175,75)(175,53){3}{3}\Gluon(175,75)(152,53){3}{4}\Gluon(175,75)(198,53){3}{4}
\Gluon(100,75)(77,53){3}{4} \Gluon(100,75)(123,53){3}{4}
\end{picture}
\end{equation}

\vspace{3.5cm}

The ghost-triangle type diagram on the left is the IR-dominant and yields the behavior  
of the $A[m,2n]$ function  in eq. (\ref{dse}).
Diagrams marked (2b) (see sketch below, left diagram) arise due 
to the four-gluon coupling, that allows two external gluon legs linked 
together. Finally we have diagrams of type (3), that we 
define as those with no external gluon legs ($m=0$) that require 
separate treatment. These are the only loop diagrams in the 
Dyson-Schwinger equations for multighost scattering kernels.
Since there is only one bare  ghost-gluon vertex in the 
Lagrangian these can all be resummed to the diagram on the right 
(that is leading and again yields our IR counting):
\begin{equation} \nonumber
\begin{picture}(-125,0)(275,100)
\Boxc(100,40)(50,26) 
\put(80,38){$m,2n$}  
\put(75,0){$\kappa(n-m+4)$}
\Gluon(100,75)(125,100){3}{4} \Gluon(100,75)(75,100){3}{4}
\Gluon(100,75)(77,53){3}{4} \Gluon(100,75)(123,53){3}{4}
\Boxc(175,40)(50,26) \DashLine(152,53)(175,75){4}
\DashLine(175,75)(175,100){4}
\Gluon(175,75)(198,53){3}{4}
\put(155,38){$m+1,2n$} \put(155,0){$\kappa (n-m)$}
\end{picture}
\end{equation}
\vspace{3.3cm}


This exhausts the possible types of proper diagrams
that can be resummed from the perturbative Yang-Mills series.

Once we have established that the ansatz eq. (\ref{result}) is a valid
solution of the Dyson-Schwinger equations in Landau gauge, we turn
to phenomenological applications. These are based in the sister Coulomb
gauge, also transverse but with a Hamiltonian formulated at equal time.
The obvious phenomenological consequence of the gluon mass gap is that 
hybrid mesons ($q\bar{q}g$), glueballs, and other states whose leading
wavefunction in Fock space involve constituent glue, will be rather massive,
in agreement with estimates in lattice gauge theory 
\cite{Morningstar:1999rf,Meyer:2003rf}. We complement lattice studies in that
our semianalytical methods in the continuum can readily be used to study
higher spin states, not easily accessible in the lattice.
Since glueballs do not have exotic quantum numbers they are the most 
difficult unconventional states to be disentangled from the meson spectrum.
However they leave some trace in  Regge phenomenology.  There is no doubt
among theorists that Regge behavior is present in QCD, but the BFKL,
BKP equations are valid only for $s>>-t>>\Lambda_{QCD}$,  whereas the 
experimental evidence so far has been gathered at small or zero $-t$ (mostly
in the form of total cross sections), therefore
non-perturbative approaches are necessary to study Regge trajectories
at low $t$. The time-honored approach to obtain the intercept of a Regge
trajectory at $t=M^2=0$ has  been to study the resonances on the positive
$(M^2,J)$  quadrant of a Chew-Frautschi plot.

We address these states by means of a model Hamiltonian employing the Coulomb
gauge formulation and fields, where the kernel is approximated
 by a charge-density to charge-density interaction  with a Cornell potential.
In a first approximation one neglects (suppressed)  transverse gluon exchange,
that can later be incorporated in perturbation theory (a model computation
along these lines is available \cite{Llanes-Estrada:2004wr}).
In the gluon sector this effective QCD Hamiltonian is
\begin{eqnarray}
H^{g}_{eff} &=&  
Tr  \int d {\bf x}\left[ {\bm \Pi}^a({\bf x})\cdot {\bm \Pi}^a({\bf x}) 
+ {\bf B}_A^a({\bf x})\cdot{\bf B}_A^a({\bf x}) \right ]  \nonumber \\
&-& \frac {1} {2}  \int d {\bf x} d {\bf y}\thinspace
\rho^a_g({\bf x}) V({\bf x},{\bf y})  \rho^a_g({\bf y}) \ ,
\end{eqnarray}
with color charge density 
$\rho^a_g({\bf x}) = f^{abc}{\bf A}^b({\bf x})\cdot{\bf {\bm \Pi}}^c({\bf x})$,
gauge fields   ${\bf A}^a$, conjugate momenta ${\bm \Pi}^a = - {\bf
E}^a$ and Abelian components
${\bf B}^a_A = {\bm \nabla} \times {\bf A}^a$,   for $a = 1,2,...8$.
The  Fourier transform to momentum space yields
\begin{eqnarray}
\label{colorfields2}
{\bf A}^a({\bf{x}}) &=&  \int
\frac{d{\bf{q}}}{(2\pi)^3}
\frac{1}{\sqrt{2\omega_k}}[{\bf a}^a({\bf{q}}) + {\bf a}^{a\dag}(-{\bf{q}})]
e^{i{\bf{q}}\cdot
{\bf {x}}} \ , \ \ \
\\ 
{\bm \Pi^a}({\bf{x}}) &=& -i \int \frac{d{\bf{q}}}{(2\pi)^3}
\sqrt{\frac{\omega_k}{2}}
[{\bf a}^a({\bf{q}})-{\bf a}^{a\dag}(-{\bf{q}})]e^{i{\bf{q}}\cdot
{\bf{x}}}  ,
\end{eqnarray}
satisfying the Coulomb gauge transverse condition, \\  
${\bf q}\cdot {\bf a}^a ({\bf q}) = \\ (-1)^\mu q_{\mu} a_{-\mu} ^a ({\bf q}) =0$.
Here $a_{\mu}^a({\bf{q}})$ ($\mu = 0, \pm 1$)
are the bare  gluon Fock operators  from which, by a
Bogoliubov-Valatin canonical transformation, the dressed gluon or quasiparticle
operators, 
$\alpha^a_{\mu} ({\bf{q}}) = \cosh \Theta(q) \,
a_{\mu}^a({\bf{q}}) +
\sinh \Theta(q) \,  a_{\mu}^{a\dagger}(-{\bf{q}})$, emerge.
This  similarity transformation is a
hyperbolic rotation similar to the BCS fermion treatment. These operators excite
constituent gluon quasiparticles from the BCS vacuum, $|\Omega>_{\rm
BCS}$,
and satisfy the  transverse commutation relations,
$[\alpha^a_{\mu}({\bf q}),\alpha^{b \dagger}_{\nu}({\bf q}')]=\delta_{ab}
(2\pi)^3 \delta^3({\bf q}-{\bf q}')D_{{\mu} {\nu}}({\bf q})  $,
with  $D_{{\mu} {\nu}}({\bf q}) = \left( 
\delta_{{\mu}{\nu}}- (-1)^{\mu}\frac{q_{\mu} q_{-\nu}}{q^2} \right)$.
Finally, the quasiparticle or gluon self-energy, 
$\omega(q) = q
e^{-2\Theta(q)}$, satisfies a gap equation  \cite{letter96glue}. A 
more sophisticated treatment is also available in the literature
\cite{Szczepaniak:2003ve},

In this report we concentrate on glueballs (hybrid mesons can be found
in \cite{Llanes-Estrada:2000hj}). 
Two-gluon glueballs \cite{Llanes-Estrada:2000jw} have
been variationally computed in the Russell-Saunders (or $L-S$) scheme.
The ground state is the $L=0$, $S=0$ scalar $J^{PC}=0^{++}$ that we
employ to fix the cutoff regulating the divergence in the gluon self-energy, 
to agree with the calculation of Morningstar and Peardon 
\cite{Morningstar:1999rf}.
The rest of the parameters being fixed from the quark sector, all other
glueball masses are predictions of the approach. 
If we plot the spectrum in a Chew-Frautschi plot to reveal Regge trajectories,
this state cannot be on the leading (left-most) trajectory since it has
total $J=0$. Its orbital excitations with $S=0$, $L=J=2,4...$ fall on
a linear Regge trajectory with slope close to the pomeron.
The second state is the $2^{++}$ in which both spins are aligned, and 
this one does fall on the pomeron Regge trajectory as observed first 
by Simonov (\cite{Kaidalov:1999yd} and references therein).  
This is natural since the intercept of a Regge trajectory with a line
of integer $J$ usually (but not always) produces a resonance in the spectrum.
However, the pomeron having a small slope (0.2-0.3 by modern fits
\cite{Pelaez:2004vs})
cannot be represented by ordinary mesons. Casimir color scaling between
the gluon-gluon and the quark-antiquark interaction provide for an enhanced
string tension in the glueball system that supports the pomeron-glueball
conjecture (the slope $\alpha '(t) \propto 1/\sigma$).
Turning now to the three body problem, the three-gluon variational 
wavefunction  is
\begin{eqnarray}
\lefteqn{\hspace{.5cm}\arrowvert \Psi^{JPC} \rangle = \int d{\bf q}_1 d{\bf q}_2
d{\bf q}_3 
\delta({\bf q}_1 + {\bf q}_2 + {\bf q}_3 )} \\ 
& & 
\hspace{-.3cm}  
F^{JPC}_{\mu_1 \mu_2 \mu_3}({\bf q}_1,{\bf q}_2,{\bf q}_3) 
C^{abc}
\alpha^{a\dagger}_{\mu_1}({\bf q}_1)
\alpha^{b\dagger}_{\mu_2}({\bf q}_2) \alpha^{c\dagger}_{\mu_3}({\bf q}_3)
\arrowvert \Omega \rangle_{\rm BCS} \ , \nonumber
\end{eqnarray}
with  summation over repeated indices. The color tensor $C^{abc}$  is 
either totally antisymmetric $f^{abc}$ (for $C =$ 1) or symmetric 
$d^{abc}$ (for $C =$ -1).
Boson statistics thus requires the $C =$ -1 oddballs to have a
symmetric space-spin wavefunction taken here to have form 
\begin{eqnarray} 
F^{JPC}_{\mu_1 \mu_2 \mu_3}({\bf q}_1,{\bf q}_2,{\bf q}_3) = [c_{12} f(q_1,q_2) +
c_{23} f(q_2,q_3) +
\nonumber \\
 c_{13} f(q_1,q_3)]
 [Y_L^{\lambda}(\hat{\bf q}_1) + Y_L^\lambda (\hat{\bf q}_2) + 
Y_L^{\lambda}(\hat{\bf q}_3)] \ ,
\label{eqf12} \\
c_{12} = \langle 1 \mu_1 1 \mu_2 \arrowvert s \mu_s \rangle
\langle s \mu_s 1 \mu_3 \arrowvert S \mu \rangle 
\langle L \lambda S\mu \arrowvert J
{\cal M}\rangle \ ,
\label{c12}
\end{eqnarray}
sufficient to analyze the lightest states.
The other two coefficients in Eq. (\ref{eqf12}) can be obtained
by permuting the indices in Eq. (\ref{c12}).
Several forms for the variational radial wavefunction, $f(q,q')$,
involving two variational parameters, $\beta$ and $\beta'$, were 
investigated including a separable form, $f(q) f(q')$, which only involves 
one parameter. 
From previous experience \cite{Llanes-Estrada:2000hj}, reliable, accurate variational
solutions can be obtained if these  functions have a
bell-shaped form with scalable variational parameters.

\begin{figure}[!thb]
\vspace*{9.0cm}
\begin{center}
\includegraphics{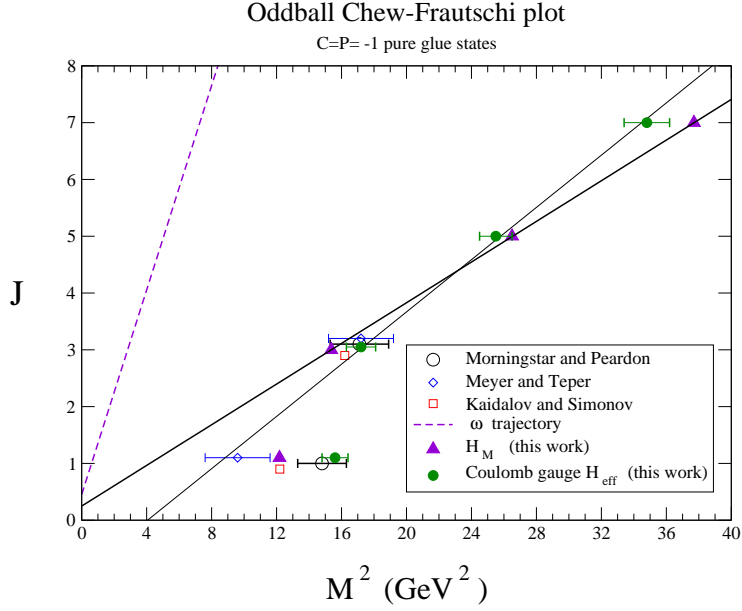}
\caption[*]{Chew-Frautschi plot for the $J^{--}$ glueball spectrum.
oddballs fall on linear Regge trajectories with a slope close to the 
pomeron,
0.25 in our preferred model $H_{eff}$. The leading trajectory starts at 
the $3^{--}$ oddball (three gluons with spins aligned, with excitations
of increasing even orbital $L$) and has very low 
intercept at $J=0$, subleading to the conventional $\omega$ Regge trajectory.
For comparison we also include existing lattice and 
constituent gluon model estimates.} 
\end{center}
\label{oddballspectrum}
\end{figure}

The $+-$ oddballs are not reported in this paper since they likely require
the use of three body forces \cite{Szczepaniak:2005xi}. The ground state of the three-gluon
system is calculated in our approach to be a $0^{-+}$ glueball (because of 
color, an annihilation diagram in \cite{Llanes-Estrada:2005jf} cannot contribute
in the odd-C sector), but this
is heavier than the corresponding $p$-wave excitation in the  two-body
system. It is interesting to note also that these $0^{-+}$ glueballs
appear in the spectrum as apparent $\eta_c$ excitations which have been
reexamined in recent B-meson factory experiments, leading to a revised
$\eta_c(2S)$ mass.
Finally we present our $(PC = --)$ spectrum in  the figure
\ref{oddballspectrum}.
For comparison we also show the good agreement with lattice computations. 
In addition we give an existing constituent gluon model calculation 
\cite{Kaidalov:1999yd} and 
another by us. The constituent model treatment of \cite{Hou:1982dy} does not report on higher spin
states and we leave it out.  One can immediately observe that  the leading
oddball Regge trajectory, in analogy with the pomeron, the odderon, passes
by the $3^{--}$ trajectory (three gluons, all spins aligned) and not
by the $1^{--}$ as erroneously assumed in the literature. We look forward
to further lattice computations for higher spin states to confirm this 
point.

Also shown in the figure is the conventional $\omega$ Regge trajectory. As can 
be seen  and was to be expected due to the smaller color factor, it has much
larger slope but higher intercept. At $t=0$ it will clearly dominate
scattering amplitudes and the odderon should not be visible, as is the 
case in low $-t$ searches at DESY \cite{desy}.
Although in our best variational model the odderon will not be observable
(by the time the two trajectories intersect,  $t=-2.3GeV^2$ and $J<-1$,
a linear extrapolation from timelike $t$ is no more valid),
in our approach,
we can not categorically preclude future searches,
either at DESY or BNL, discovering an odderon signature.
Such searches, however, should also be conducted at suffuciently higher -t
(greater than $.5\ GeV^2$).
Nevertheless we remain pessimistic about the prospects
for observing the odderon especially since a recent extension by Kaidalov 
and Simonov of their approach  to higher spin states 
concurs with our findings  \cite{kaidalov2}.

In concluding we note this is a period of timely
opportunities for exciting phenomenology related to dynamical gluon
mass generation and propagator suppression.
For example, the recent CLEO  measurement \cite{Andreotti:2005vu} of the
$1^{+-}$ meson mass, almost degenerate with the $1^{++}$, shows that the 
hyperfine interaction is quite short range.  This has traditionally been 
a conundrum
of the quark model. If a long range Coulomb form is supposed for the physical
gluon exchange and $\Delta V(r) \propto -\delta(r)$ the hyperfine potential has
to be treated in perturbation theory (or the wavefunction falls to the origin)
Then smearing parameters have to be introduced whose meaning is unclear.
In contrast, the modern, field-theory based formulation cleanly separates
the Cornell potential from the transverse exchange. This can be suppressed, its
range being of order the inverse gluon gap $1/E_{min}(k)$, and the hyperfine
interaction does not need to be so large as to provide $\pi$-$\rho$ splitting,
since chiral symmetry accomplishes this \cite{Bicudo:1989si} and the transverse
interaction can be treated in perturbation theory.

\section*{Acknowledgements} F. J. L. thanks the organizers of this
interesting meeting at the 30th BFKL birthday, where QCD, Regge theory 
and experimental data where harmoniously  combined, for their warm 
hospitality and the opportunity to report on this work.

\end{document}